\documentclass[jcp,aip,amsmath,amssymb,reprint,superscriptaddress]{revtex4-2}

\usepackage[dvipdfmx]{graphicx}
\usepackage{color}
\usepackage{epsf}
\usepackage{bm}
\usepackage{physics}
\usepackage{lipsum}
\usepackage[normalem]{ulem}
\usepackage{txfonts}
\usepackage{booktabs}
\usepackage{multirow}
\usepackage{here}

\begin{document}

%%%%%%%%%%%%%%%%%%%%%%%%%%%%%%%%%%%%%%%%%%%%%%%%%%%%%%%%%%%%%%%%%%%%%%%%%%%%%%%%%%%%%%%%%%%%%%%%%%%%%%%%%%%%%%%%%%%%%
\title{Unveiling interatomic distances influencing the reaction
 coordinates in alanine dipeptide isomerization:
 An explainable deep learning approach}

%%%%%%%%%%%%%%%%%%%%%%%%%%%%%%%%%%%%%%%%%%%%%%%%%%%%%%%%%%%%%%%%%%%%%%%%%%%%%%%%%%%%%%%%%%%%%%%%%%%%%%%%%%%%%%%%%%%%%
\author{Kazushi Okada}
\affiliation{Division of Chemical Engineering, Department of Materials Engineering Science, Graduate School of Engineering Science, Osaka University, Toyonaka, Osaka 560-8531, Japan}

\author{Takuma Kikutsuji}
\affiliation{Division of Chemical Engineering, Department of Materials Engineering Science, Graduate School of Engineering Science, Osaka University, Toyonaka, Osaka 560-8531, Japan}

\author{Kei-ichi Okazaki}
\email{keokazaki@ims.ac.jp}
\affiliation{Research Center for Computational Science, Institute for Molecular Science, Okazaki, Aichi 444-8585, Japan}
\affiliation{Graduate Institute for Advanced Studies, SOKENDAI, Okazaki,
Aichi 444-8585, Japan}

\author{Toshifumi Mori}
\email{toshi\_mori@cm.kyushu-u.ac.jp}
\affiliation{Institute for Materials Chemistry and
Engineering, Kyushu University, Kasuga, Fukuoka 816-8580, Japan}
\affiliation{Interdisciplinary Graduate School of Engineering Sciences,
Kyushu University, Kasuga, Fukuoka 816-8580, Japan}

\author{Kang Kim}
\email{kk@cheng.es.osaka-u.ac.jp}
\affiliation{Division of Chemical Engineering, Department of Materials Engineering Science, Graduate School of Engineering Science, Osaka University, Toyonaka, Osaka 560-8531, Japan}

\author{Nobuyuki Matubayasi}
\email{nobuyuki@cheng.es.osaka-u.ac.jp}
\affiliation{Division of Chemical Engineering, Department of Materials Engineering Science, Graduate School of Engineering Science, Osaka University, Toyonaka, Osaka 560-8531, Japan}

%%%%%%%%%%%%%%%%%%%%%%%%%%%%%%%%%%%%%%%%%%%%%%%%%%%%%%%%%%%%%%%%%%%%%%%%%%%%%%%%%%%%%%%%%%%%%%%%%%%%%%%%%%%%%%%%%%%%%
\date{\today}

%%%%%%%%%%%%%%%%%%%%%%%%%%%%%%%%%%%%%%%%%%%%%%%%%%%%%%%%%%%%%%%%%%%%%%%%%%%%%%%%%%%%%%%%%%%%%%%%%%%%%%%%%%%%%%%%%%%%%
\begin{abstract}
The present work shows that the free energy landscape associated
 with 
alanine dipeptide isomerization can be effectively represented by specific
 interatomic distances without explicit reference to dihedral angles. 
Conventionally, two stable states of alanine dipeptide in vacuum,
 \textit{i.e.}, C$7_{\mathrm{eq}}$ ($\beta$-sheet structure) and
 C$7_{\mathrm{ax}}$ (left handed $\alpha$-helix structure), have been
 primarily characterized using the main chain
 dihedral angles, $\varphi$ (C-N-C$_\alpha$-C) and
$\psi$ (N-C$_\alpha$-C-N).
However, our recent deep learning combined with ``Explainable AI'' (XAI)
 framework has shown that
 the transition state can be adequately captured by a free energy landscape using
 $\varphi$ and $\theta$ (O-C-N-C$_\alpha$) [T. Kikutsuji, \textit{et al.}
 \textit{J. Chem. Phys.} \textbf{156}, 154108 (2022)].
In perspective of extending these insights to other collective
 variables, 
a more detailed characterization of transition state is required.
In this work, we employ the interatomic
 distances and bond angles as input variables for deep learning, rather than the
 conventional and more elaborate dihedral angles.
Our approach utilizes deep learning to investigate whether changes in
 the main chain dihedral angle can be expressed in terms of interatomic
 distances and bond angles.
Furthermore, by incorporating XAI into our predictive analysis, 
we quantified the importance of each input variable
 and succeeded in clarifying the specific interatomic distance that affects the
 transition state. 
The results indicate that constructing a free energy landscape based on 
 using the identified interatomic distance can clearly distinguish between
 the two stable states and provide a comprehensive explanation for the
 energy barrier crossing.
\end{abstract}
%%%%%%%%%%%%%%%%%%%%%%%%%%%%%%%%%%%%%%%%%%%%%%%%%%%%%%%%%%%%%%%%%%%%%%%%%%%%%%%%%%%%%%%%%%%%%%%%%%%%%%%%%%%%%%%%%%%%%
\maketitle
%%%%%%%%%%%%%%%%%%%%%%%%%%%%%%%%%%%%%%%%%%%%%%%%%%%%%%%%%%%%%%%%%%%%%%%%%%%%%%%%%%%%%%%%%%%%%%%%%%%%%%%%%%%%%%%%%%%%%
\section{Introduction}

In various complex molecular systems, such as protein conformational changes,
the central task often involves the selection of one or two
collective variables (CVs) 
to characterize the free energy landscape (FEL).
Specifically, 
the probability distribution function
$P(r)$ for a chosen CV $r$ from a large number of CV candidates is obtained
through molecular dynamics (MD) simulations.
By taking its logarithmic form, $F(r)=-k_\mathrm{B} T \ln P(r)$
serves as the representation of the FEL.~\cite{zuckerman2010Statistical}
Within the FEL, if stable states are distinguished by a saddle
point, and moreover the transition pathway passes through the saddle point
on the FEL, the variable $r$ is regarded as the reaction
coordinate (RC) governing the target 
change.~\cite{peters2017Reaction}
In this context, the saddle point of the FEL can be regarded as the transition state (TS).

\begin{figure*}[t]
\includegraphics[width=0.9\textwidth]{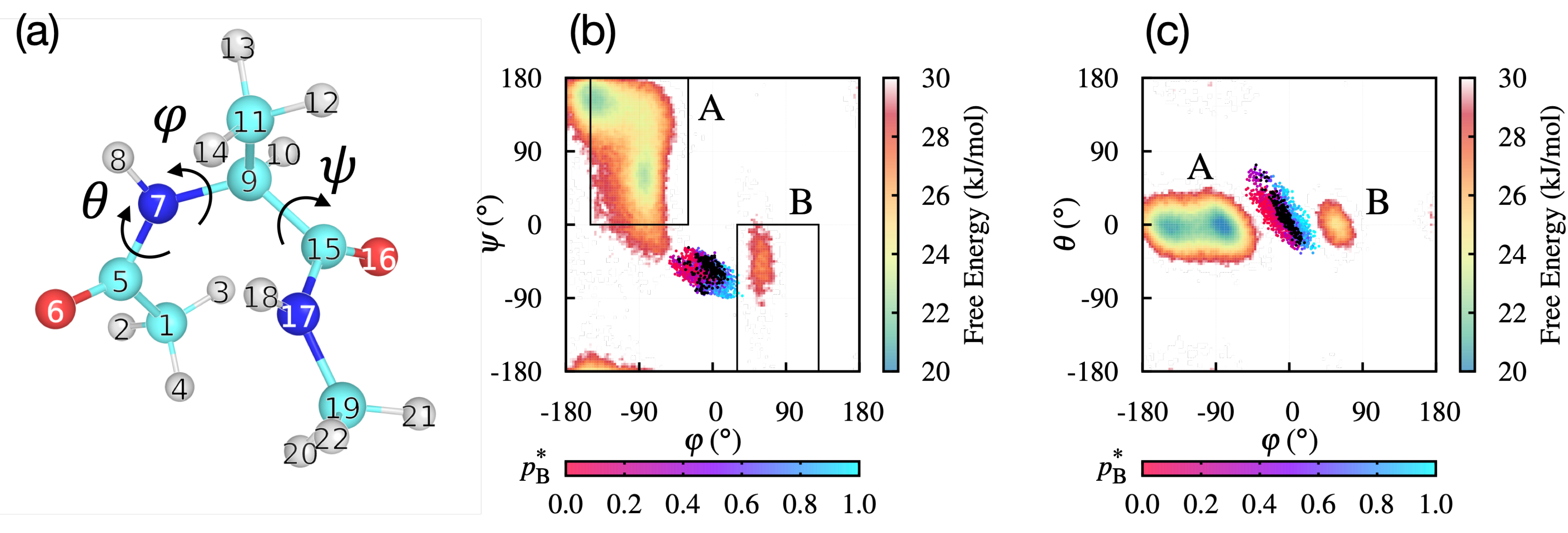}
\caption{
(a) Index assigned to alanine dipeptide atoms.
Major three dihedral angles $\varphi$, $\psi$, and $\theta$ are also described.
(b) and (c): 
Two-dimensional FELs using ($\varphi$, $\psi$) (b) and ($\varphi$,
 $\theta$) (c).
In (b), the black boxes describe states A [$( -150^\circ, 0^\circ ) \le
 ( \varphi, \psi ) \le (
-30^\circ, 180^\circ )$] and B [$( 30^\circ,-180^\circ ) \le ( \varphi, \psi ) \le (
130^\circ, 0^\circ )$].
The points represents sampled 1,000 shooting points (training 
 data set), which are colored by
 $p_\mathrm{B}^*$ values given in the bottom color bar.
Note that the total number of shooting points is 2000 and that 1000 of
 them are shown in (b) and (c) as the training dataset.
In addition, the points with $p_\mathrm{B}^* \sim 0.5$ $(0.45 \le p_\mathrm{B}^* \le 0.55)$ are marked in black dots.
}
\label{fig:ala2}
\end{figure*}

The isomerization of alanine dipeptide serves as 
a model for dihedral angle changes in proteins and represents 
a benchmark in exploring the complexities of FEL (see Fig.~\ref{fig:ala2}(a)).
Conventionally, the FEL of the alanine dipeptide isomerization has been described as a function of
two key dihedral angles, $\varphi$ (C-N-C$_\alpha$-C) and
$\psi$ (N-C$_\alpha$-C-N), often referred to as the
Ramachandran plot.~\cite{ramachandran1968Conformation} 
In vacuum, two stable states, C$7_{\mathrm{eq}}$ ($\beta$-sheet
structure) and C$7_{\mathrm{ax}}$ (left handed $\alpha$-helix structure),
are well 
characterized by the FEL using $\varphi$ and $\psi$.
Hereafter, C$7_{\mathrm{eq}}$ and C$7_{\mathrm{ax}}$ are denoted as
states A and B, respectively (see Fig.~\ref{fig:ala2}(b)).

The transition path sampling and committor analysis, when applied to the
alanine dipeptide in vacuum, 
have unveiled the necessity of incorporating another dihedral angle,
$\theta$ (O-C-N-C$_\alpha$), to adequately describe the FEL for the
conformational change between the two states A and B.~\cite{bolhuis2000Reaction}
Here, the committor, denoted as $p_\mathrm{B}^*(\bm{x})$,
represents the probability of
the trajectories reaching state B before state A
starting from the
initial configuration $\bm{x}$ produced by the Maxwell--Boltzmann
velocity distribution at temperature $T$.~\cite{du1998Transition, geissler1999Kinetic,
bolhuis2002Transition, hummer2004Transition, best2005Reaction, e2005Transition,
peters2006Using, peters2006Obtaining, peters2007Extensions, peters2010TP, peters2016Reaction, jung2017Transition}
The selected CVs used to describe the FEL can be considered the primary
contributors to the RC 
if the distribution of evaluated committor $p_\mathrm{B}^*(\bm{x})$ of
varieties of configurations $\bm{x}$ continuously
changes from 0 to 1.~\cite{hagan2003Atomistic, 
pan2004Dynamics, rhee2005OneDimensional, berezhkovskii2005Onedimensional,
moroni2005Interplay, branduardi2007Free,
quaytman2007Reaction, beckham2007SurfaceMediated, peters2008Path,
beckham2008Evidence, antoniou2009Stochastic, xi2013Hopping,
jungblut2013Optimising, 
barnes2014Reactiona, mullen2014Transmission, mullen2015Easy, 
lupi2016Preordering, 
ernst2017Identification, diazleines2018Maximum, joswiak2018Ion,
okazaki2019Mechanism, mori2020Dissecting,
schwierz2020Kinetic, silveira2021Transitiona}
Correspondingly, the TS is characterized by the collection of configurations $\bm{x}$
exhibiting $p_\mathrm{B}^* =0.5$.

%%%%%%%%%%%%%%%%%%%%%%%%%%%%%%%%%%%%%%%%%%%%%%%%%%%%%%%%%%%%%%%%%%%%%%%%%%%%%%%%%%%%%%%%%%
%%%%%%%%% Fig. 1 %%%%%%%%%%%%%%%%%%%%%%%%%%%%%%%%%%%%%%%%%%%%%%%%%%%%%%%%%%%%%%%%%%%%%%%%%
%%%%%%%%%%%%%%%%%%%%%%%%%%%%%%%%%%%%%%%%%%%%%%%%%%%%%%%%%%%%%%%%%%%%%%%%%%%%%%%%%%%%%%%%%%

%%%%%%%%%%%%%%%%%%%%%%%%%%%%%%%%%%%%%%%%%%%%%%%%%%%%%%%%%%%%%%%%%%%%%%%%%%%%%%%%%%%%%%%%%%
%%%%%%%%% Fig. 2 %%%%%%%%%%%%%%%%%%%%%%%%%%%%%%%%%%%%%%%%%%%%%%%%%%%%%%%%%%%%%%%%%%%%%%%%%
\begin{figure}[t]
\includegraphics[width=0.4\textwidth]{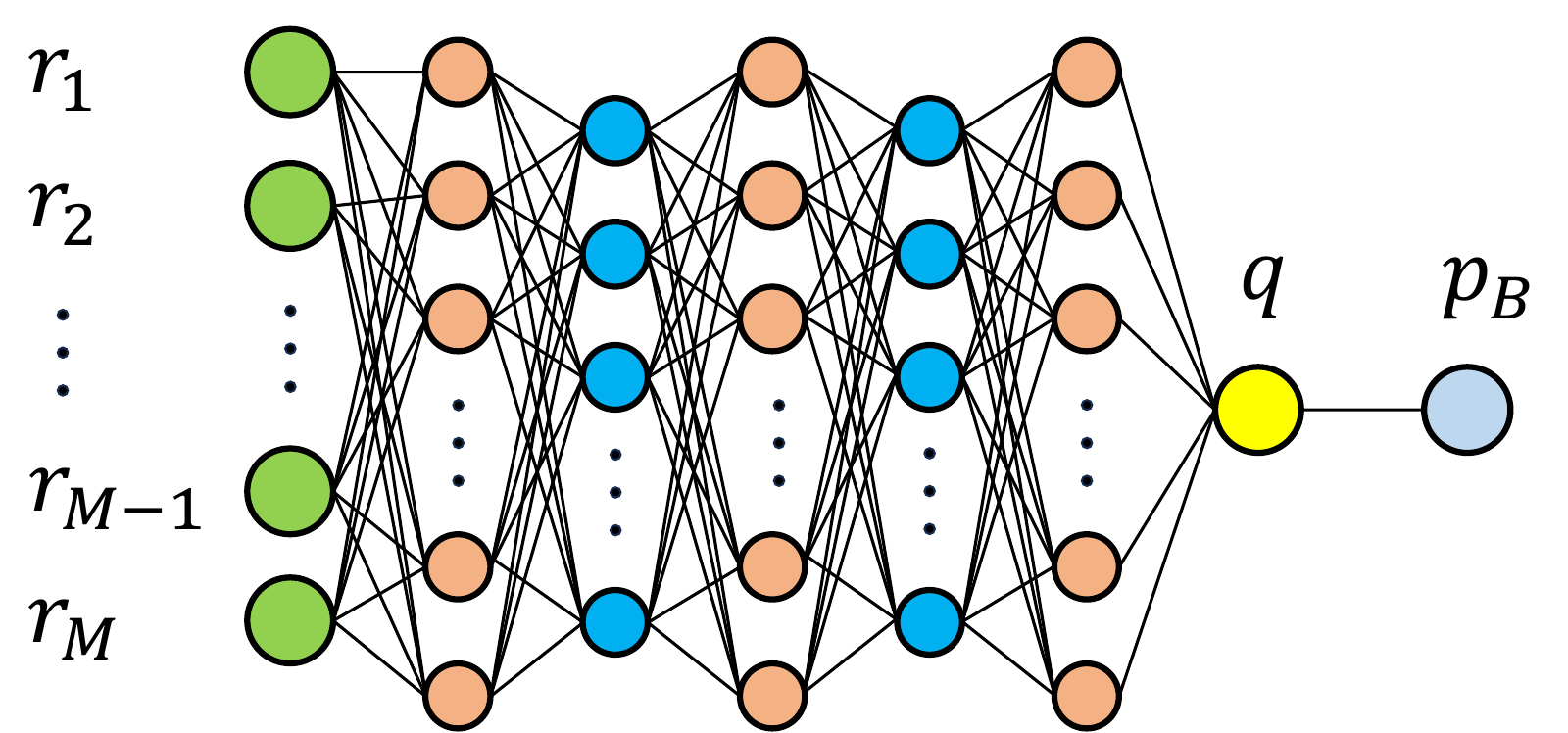}
\caption{
Schematic diagram of the neural network architecture for the training of the relationship between
committor value
 $p_\mathrm{B}^*$ and $M$ CVs plus 1 bias term $\bm{r}=(1, r_1, r_2, \cdots, r_M)$, and predicting committor
 $p_\mathrm{B}$ as a sigmoidal committor function
$p_\mathrm{B}(q)=(1+\tanh(q))/2$. 
The neural network involves five hidden layers, of which the odd- and even-numbered layers had 400
and 200 nodes, respectively.
Note that the node representing the bias term is omitted from the diagram.}
\label{fig:DNN}
\end{figure}
%%%%%%%%%%%%%%%%%%%%%%%%%%%%%%%%%%%%%%%%%%%%%%%%%%%%%%%%%%%%%%%%%%%%%%%%%%%%%%%%%%%%%%%%%%

Machine learning stands out as one of the most promising methods for
automatically identifying the appropriate CVs relevant to
RC.~\cite{sittel2016Robust, 
sultan2018Automated, wehmeyer2018Timelagged, mardt2018VAMPnets,
bittracher2018Datadriven, chen2018Molecular, ribeiro2018Reweighted, 
rogal2019NeuralNetworkBased, wang2020Machine, bonati2020DataDriven, sidky2020Machine, wang2021State,
zhang2021Deep, frassek2021Extended, hooft2021Discovering, 
bonati2021Deepa, chen2021Collective, belkacemi2022Chasing, neumann2022Artificial, 
ray2023Deepa, jung2023Machineguided, bonati2023Unified,
singh2023Variationala, liang2023Probinga, lazzeri2023Molecular}
The pioneering work by Ma and Dinner introduced an
automated search approach for identifying adequate CVs in alanine dipeptide
isomerization using
a genetic algorithm.~\cite{ma2005Automatic}
Expanding on this, we have utilized alternative machine learning
techniques such as linear regression and deep neural networks, considering all 
possible dihedral angles of alanine dipeptide as candidate
CVs.~\cite{mori2020Learning, kikutsuji2022Explaining}
Significantly, our investigation highlighted the effectiveness of the ``Explainable AI
(XAI)'' method in providing a local
explanation model for the data along the RC.
As a result, the description of 
the separatrix line on the FEL has been suitably characterized 
by $p_\mathrm{B}^* =0.5$, employing two dihedral angles
$\varphi$ and $\theta$ (see Fig.~\ref{fig:ala2}(c)).~\cite{kikutsuji2022Explaining}
Recently, 
sophisticated methodologies, such as persistent
homology~\cite{manuchehrfar2021Exacta} and Light Gradient Boosting
Machine (LightGBM),~\cite{naleem2023Exploration} have been utilized in transition path samplings for the
isomerization 
of alanine dipeptide.

It is of significance to emphasize that  
the scope of CVs extends beyond dihedral angles to
include interatomic distances and bond angles, all of which represent 
plausible candidates for describing 
the FEL.
Dihedral angles, derived from four adjacent atoms, play a crucial role
in elucidating protein conformations, while interatomic distances and
bond angles, being variables of relatively simpler computation, provide a
more general means of analysis compared to dihedral angles.
Hence, in this study, we employ interatomic distances and bond
angles of alanine dipeptide as input variables for the neural network. 
Furthermore, the application of XAI facilitates a
comprehensive understanding of CVs 
that influence the RC, without relying solely on dihedral angles.

\section{Methods}

%%%%%%%%%%%%%%%%%%%%%%%%%%%%%%%%%%%%%%%%%%%%%%%%%%%%%%%%%%%%%%%%%%%%%%%%%%%%%%%%%%%%%%%%%%
%%%%%%%%% Fig. 3 %%%%%%%%%%%%%%%%%%%%%%%%%%%%%%%%%%%%%%%%%%%%%%%%%%%%%%%%%%%%%%%%%%%%%%%%%
\begin{figure*}[t]
\includegraphics[width=0.9\textwidth]{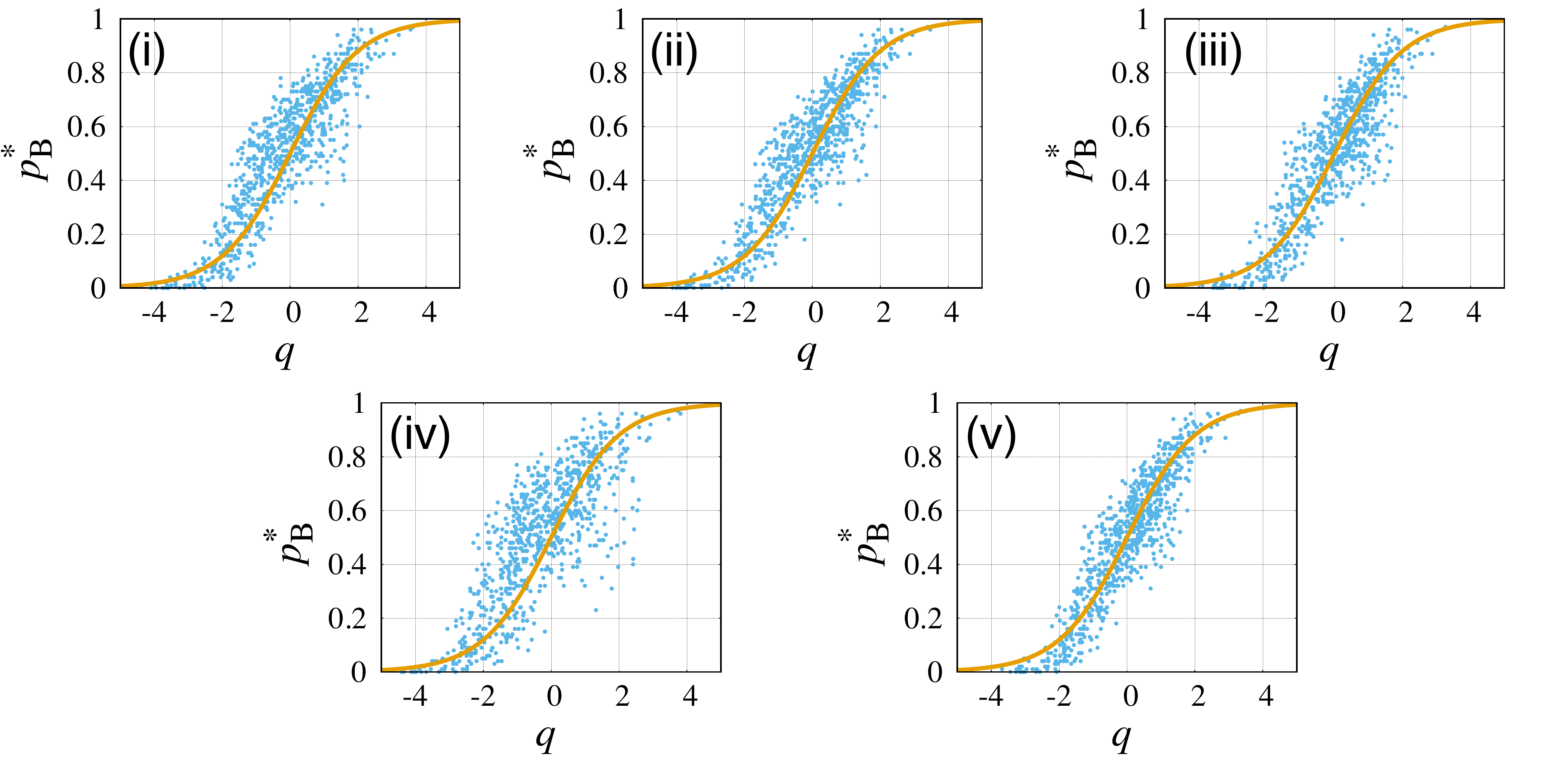}
\caption{
Relationship between $p_\mathrm{B}^*$ and $q$ derived from the neural
 network trained model using test dataset (800 points).
The orange curve represents the sigmoidal function
$p_\mathrm{B}(q)=(1+\tanh(q))/2$.
Input variables for each case are as follows: (i) 90 dihedral angles,
 210 interatomic distances, and 36 bond angles, 
(ii) 90 dihedral angles and 210 interatomic distances, 
(iii) 210 interatomic distances and 36 bond angles, 
(iv) 90 dihedral angles and 36 bond angles, 
and (v) 210 interatomic distances.
}
\label{fig:pB_q}
\end{figure*}
%%%%%%%%%%%%%%%%%%%%%%%%%%%%%%%%%%%%%%%%%%%%%%%%%%%%%%%%%%%%%%%%%%%%%%%%%%%%%%%%%%%%%%%%%%

%%%%%%%%%%%%%%%%%%%%%%%%%%%%%%%%%%%%%%%%%%%%%%%%%%%%%%%%%%%%%%%%%%%%%%%%%%%%%%%%%%%%%%%%%%
%%%%%%%%% Table 1 %%%%%%%%%%%%%%%%%%%%%%%%%%%%%%%%%%%%%%%%%%%%%%%%%%%%%%%%%%%%%%%%%%%%%%%%%
\begin{table*}[t]
\caption{Top five dominant contributions and those absolute values obtained
 using LIME and SHAP for various combinations of input variable types.
The dihedral angles defined from atom numbers (X-5-7-Y) and (X-7-9-Y)
 denote $\theta$ and $\varphi$, respectively, with arbitrary atoms X and Y.
The symbol $^*$ represents sine form of the improper dihedral angle
 defined by atom numbers (1-7-5-6).
}
\centering
\begin{tabular}{l c c c c c c c c c c}
\toprule
\midrule
 \multirow{1}{*}{input variables} & \multicolumn{4}{c}{LIME}& \hspace{1cm}  & \multicolumn{4}{c}{SHAP}\\
\cmidrule(lr){2-5} 
\cmidrule(lr){7-10} 
& type & index & feature & value &  & type & index & feature & value\\
\midrule
\multirow{1}{*}{(i) dihedral angles + interatomic distances + bond angles} & 
distance & 93 & $r_{6-10}$ & 0.942 && distance & 95 & $r_{6-12}$ & 0.026\\
& distance & 95 & $r_{6-12}$ & 0.920 && distance & 94 & $r_{6-11}$ & 0.026\\
& distance & 101 & $r_{6-18}$ & 0.811 && distance & 93 & $r_{6-10}$ & 0.023\\
& distance & 100 & $r_{6-17}$ & 0.683 && dihedral angle & 87 & * & 0.020\\
& distance & 99 & $r_{6-16}$ & 0.667 && distance & 101 & $r_{6-18}$ & 0.015\\

\midrule
\multirow{1}{*}{(ii) dihedral angles + interatomic distances} & 
distance & 93 & $r_{6-10}$ & 0.948 && distance & 93 & $r_{6-10}$ & 0.033\\
& distance & 95 & $r_{6-12}$ & 0.928 && distance & 95 & $r_{6-12}$ & 0.024\\
& distance & 94 & $r_{6-11}$ & 0.769 && distance & 94 & $r_{6-11}$ & 0.020\\
& distance & 100 & $r_{6-17}$ & 0.721 && dihedral angle & 57 & $\sin\varphi$ & 0.019\\
& dihedral angle & 58 & $\sin\varphi$ & 0.704 && distance & 100 & $r_{6-17}$ & 0.019\\

\midrule
\multirow{1}{*}{(iii) interatomic distances + bond angles} & 
distance & 93 & $r_{6-10}$ & 1.691 && distance & 93 & $r_{6-10}$ & 0.039\\
& distance & 95 & $r_{6-12}$ & 1.440 && distance & 95 & $r_{6-12}$ & 0.036\\
& distance & 94 & $r_{6-11}$ & 1.100 && distance & 94 & $r_{6-11}$ & 0.032\\
& distance & 100 & $r_{6-17}$ & 1.034 && distance & 99 & $r_{6-16}$ & 0.021\\
& distance & 97 & $r_{6-14}$ & 0.801 && distance & 100 & $r_{6-17}$ & 0.017\\

\midrule
\multirow{1}{*}{(iv) diheral angles + bond angles} & 
dihedral angle & 55 & $\sin\theta$ & 2.514 && dihedral angle & 53 & $\sin\theta$ & 0.075\\
& dihedral angle & 57 & $\sin\varphi$ & 1.921 && dihedral angle & 55 & $\sin\theta$ & 0.074\\
& dihedral angle & 58 & $\sin\varphi$ & 1.636 && dihedral angle & 57 & $\sin\varphi$ & 0.056\\
& dihedral angle & 56 & $\sin\varphi$ & 1.304 && dihedral angle & 11 & $\cos\varphi$ & 0.047\\
& dihedral angle & 87 & * & 0.762 && dihedral angle & 58 & $\sin\varphi$ & 0.046\\

\midrule
\multirow{1}{*}{(v) interactomic distances} & 
distance & 93 & $r_{6-10}$ & 1.609 && distance & 93 & $r_{6-10}$ & 0.075\\
& distance & 95 & $r_{6-12}$ & 1.553 && distance & 94 & $r_{6-11}$ & 0.074\\
& distance & 94 & $r_{6-11}$ & 1.248 && distance & 99 & $r_{6-16}$ & 0.056\\
& distance & 100 & $r_{6-17}$ & 0.972 && distance & 95 & $r_{6-12}$ & 0.047\\
& distance & 101 & $r_{6-18}$ & 0.950 && distance & 101 & $r_{6-18}$ & 0.046\\

\midrule
\bottomrule
\end{tabular}
\label{table:LIME_SHAP}
\end{table*}
%%%%%%%%%%%%%%%%%%%%%%%%%%%%%%%%%%%%%%%%%%%%%%%%%%%%%%%%%%%%%%%%%%%%%%%%%%%%%%%%%%%%%%%%%%

\subsection{Simulation details}

We conducted a numerical examination of the alanine dipeptide
isomerization in 
vacuum using MD simulations. 
The system consists of a single alanine dipeptide molecule, and the MD
procedures are described in Refs.~\onlinecite{mori2020Learning,
kikutsuji2022Explaining}.
Configurations of the alanine dipeptide were collected through a
transition path sampling technique known as aimless
shooting.~\cite{peters2006Obtaining}
In this
method, trajectories are produced with distinctly sampled momenta from the
Maxwell--Boltzmann distribution at 300 K for each configuration.
Specifically, 
the aimless shooting was initiated from a configuration randomly chosen
from the region between the two states, A and B, as described in Fig.~\ref{fig:ala2}(b).
A total of 
2,000 shooting points were sampled.
Furthermore, for each shooting point, we
determined the committor $p_\mathrm{B}^*$ by running 1 ps MD simulations
100 times, employing 
random velocities from the Maxwell--Boltzmann distribution at 300 K
(additional numerical conditions can be found
in Ref.~\onlinecite{mori2020Learning}). 
We also computed all 45 dihedral angles (including 4 improper dihedral
angles), 210 interatomic
distances, and 
36 bond angles within the molecule as CV candidates for each configuration.
See Tables~S1-S3 of the supplementary material 
for detailed definitions of the dihedral angles, interatomic
distances, and bond angles.
For describing FELs in this study, 
the replica-exhange MD simulations were employed, the detail of which 
is also available in Ref.~\onlinecite{mori2020Learning}.

\subsection{Neural network learning and XAI}
\label{sec:XAI}

We employed neural network learning to predict 
the relationship between the committor distribution $p_\mathrm{B}^*$
and candidate CVs.
The architecture of the neural network is identical to 
that utilized in our previous study, as described in Fig.~\ref{fig:DNN}.~\cite{kikutsuji2022Explaining}
Bond angles are presented in the cosine form, 
while dihedral angles are transformed into both cosine and sine forms,
resulting in a total of 90 CVs.
All CVs, including dihedral angles, interatomic distances, and bond angles, were standardized.

We utilized various combinations of input variable types for the neural
networks: 
(i) 90 dihedral angles, 210 interatomic distances, and 36 bond angles, 
(ii) 90 dihedral angles and 210 interatomic distances, 
(iii) 210 interatomic distances and 36 bond angles, 
(iv) 90 dihedral angles and 36 bond angles, 
and (v) 210 interatomic distances.
Correspondingly, the total number of input variables $M$ is (i)
$M=336$, (ii) $M=300$, (iii) $M=246$, (iv) $M=126$, and (v) $M=210$, respectively.
Through the network with five hidden layers, where the odd- and even-numbered layers
consist of 400 and 200 nodes each, respectively, 
a one-dimensional variable $q$ is derived.
The neural network is trained to ensure the regression of the relationship between $q$ and
$p_\mathrm{B}^*$ to a 
sigmoidal function, represented as 
$p_\mathrm{B}(q)=(1+\tanh(q))/2$, where $q$ serves as the RC.
The dataset of CVs and committor values from 2,000 coonfigurations was partitioned into
training, validation, and test datasets at a ratio of 5:1:4.
(additional numerical conditions are
described in Ref.~\onlinecite{kikutsuji2022Explaining}). 
Note that we attempted to train the same neural network with 36 bond angle
as input variables.
However, the obtained training results were insufficient, and 
we decided not to include them in this paper.
In fact, the distribution of any bond angles fluctuates around the mean value
and is unsuitable for distinguishing between the two stable states, A
and B.

For elucidating 
the deep learning model, we employed two types of XAI
method, Local Interpretable Model-agnostic Explanation (LIME)~\cite{ribeiro2016Whya} and
Shapley Additive exPlanations (SHAP).~\cite{lundberg2017unified}
Both methodologies are designed to quantify the importance of features in deep learning
predictions and provide explanation models.
LIME provides a 
linear regression model, elucidating 
the local behavior of a target instance explained by the perturbation of
input variables with the Gauss kernel function.
Note that LIME ignores the correlation among the features, causing
the characteristics which may differ from the original dataset.
In contrast, SHAP employs an additive feature attribution
method, ensuring a fair distribution of predictions among input features
in accordance with the game-theory-based Shapley value.
Specifically, we utilized the Kernel SHAP, which
calculates the Sharpley value based on the LIME framework.
Therefore, SHAP is expected to quantify the contribution of each feature more accurately than LIME.

In a previous work, we demonstrated the efficacy of XAI to
identify the important features for the transition 
through the TS.~\cite{kikutsuji2022Explaining}
More precisely, 
the relevant CVs contributing to the RC $q$ were identified as 
$\varphi$ and
$\theta$, in contrast to $\varphi$ and $\psi$.
Note that 
an alternative application of SHAP to a deep learning using
LightGBM in the alanine dipeptide isomerization has recently been
reported.~\cite{naleem2023Exploration}
In the current study, using LIME and SHAP, we quantified the contribution of input
variables to the RC $q$ for the five different results from deep learning,
achieved by changing the 
combinations of CV types, \textit{i.e.}, dihedral angle, interatomic
distance, and bond angle.

\section{Results and Discussion}

\subsection{Learning of committor $p_\mathrm{B}^*$ and RC $q$}
\label{sec:committor_prediction}

Figure~\ref{fig:pB_q} shows the test data (800 points) results depicting the relationship
between $p_\mathrm{B}^*$ and $q$ derived from the training of the neural
networks.
Two different model evaluation metrics were employed: 
the
coefficient of determination, $R^2$, and the root mean square
error, RMSE.
Both metrics assess the fit of outcomes $q$ to the
sigmoidal function
$p_\mathrm{B}(q)=(1+\tanh(q))/2$.
The results for each case are as follows: $(R^2, \mathrm{RMSE})= (0.880, 0.130)$ (i), 
(0.899, 0.118) (ii), (0.885, 0.117) (iii), (0.889, 0.179) (iv), and (0.903, 0.103) (v).
These results signify the overall success of $p_\mathrm{B}$ prediction
through $q$.
The inclusion of bond angles as input variables leads to
a slight decrease in accuracy.
As noted in Sec.~\ref{sec:XAI}, bonding angles prove insufficient for effectively
distinguishing between the two states, 
contributing insignificantly to the outcomes of $q$.

%%%%%%%%%%%%%%%%%%%%%%%%%%%%%%%%%%%%%%%%%%%%%%%%%%%%%%%%%%%%%%%%%%%%%%%%%%%%%%%%%%%%%%%%%%
%%%%%%%%% Fig. 4 %%%%%%%%%%%%%%%%%%%%%%%%%%%%%%%%%%%%%%%%%%%%%%%%%%%%%%%%%%%%%%%%%%%%%%%%%
\begin{figure*}[t]
\includegraphics[width=0.95\textwidth]{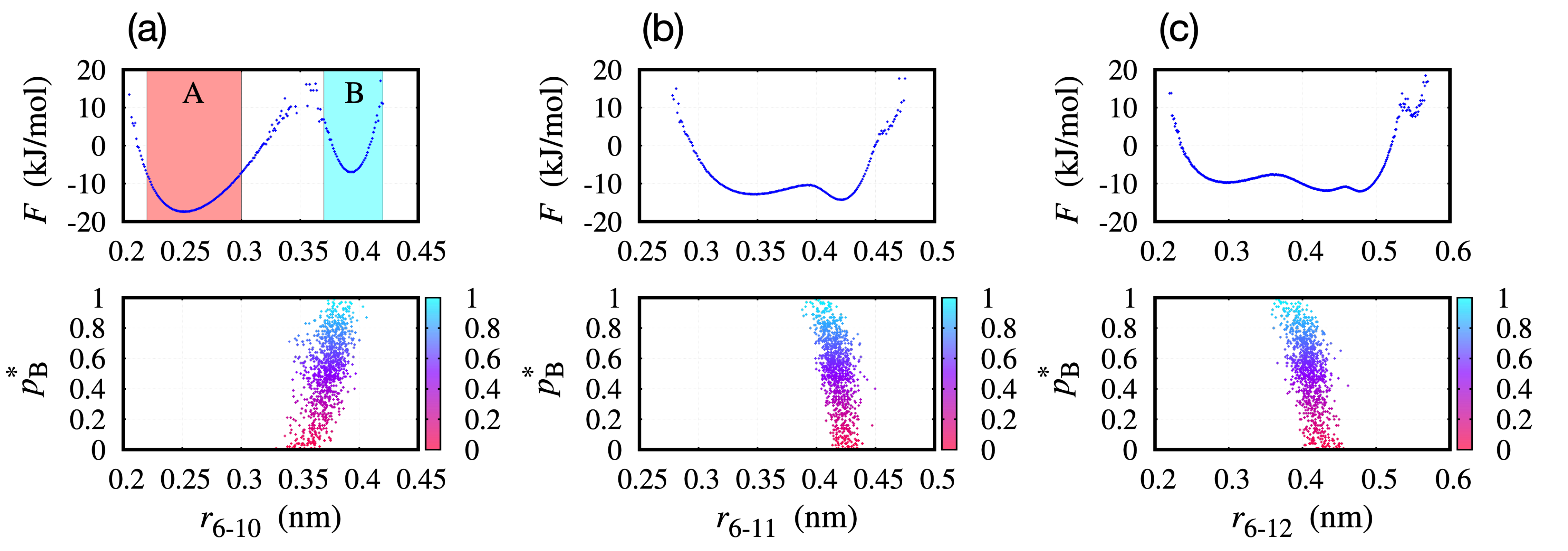}
\caption{
(Top) FEL using CV, 
$r_{6-10}$ (a), $r_{6-11}$ (b), and $r_{6-12}$ (c).
In (a), the ranges of 0.22 nm $\le$ $r_{6-10}$ $\le$
0.30 nm and 0.37 nm $\le$ $r_{6-10}$ $\le$ 0.42 nm effectively
represent states A and B, respectively.
(Bottom) Corresponding committor $p_\mathrm{B}^*$ distribution as a
 function of each CV.
}
\label{fig:PMF}
\end{figure*}
%%%%%%%%%%%%%%%%%%%%%%%%%%%%%%%%%%%%%%%%%%%%%%%%%%%%%%%%%%%%%%%%%%%%%%%%%%%%%%%%%%%%%%%%%%

%%%%%%%%%%%%%%%%%%%%%%%%%%%%%%%%%%%%%%%%%%%%%%%%%%%%%%%%%%%%%%%%%%%%%%%%%%%%%%%%%%%%%%%%%%
%%%%%%%%% Fig. 5 %%%%%%%%%%%%%%%%%%%%%%%%%%%%%%%%%%%%%%%%%%%%%%%%%%%%%%%%%%%%%%%%%%%%%%%%%
\begin{figure*}[t]
\includegraphics[width=0.8\textwidth]{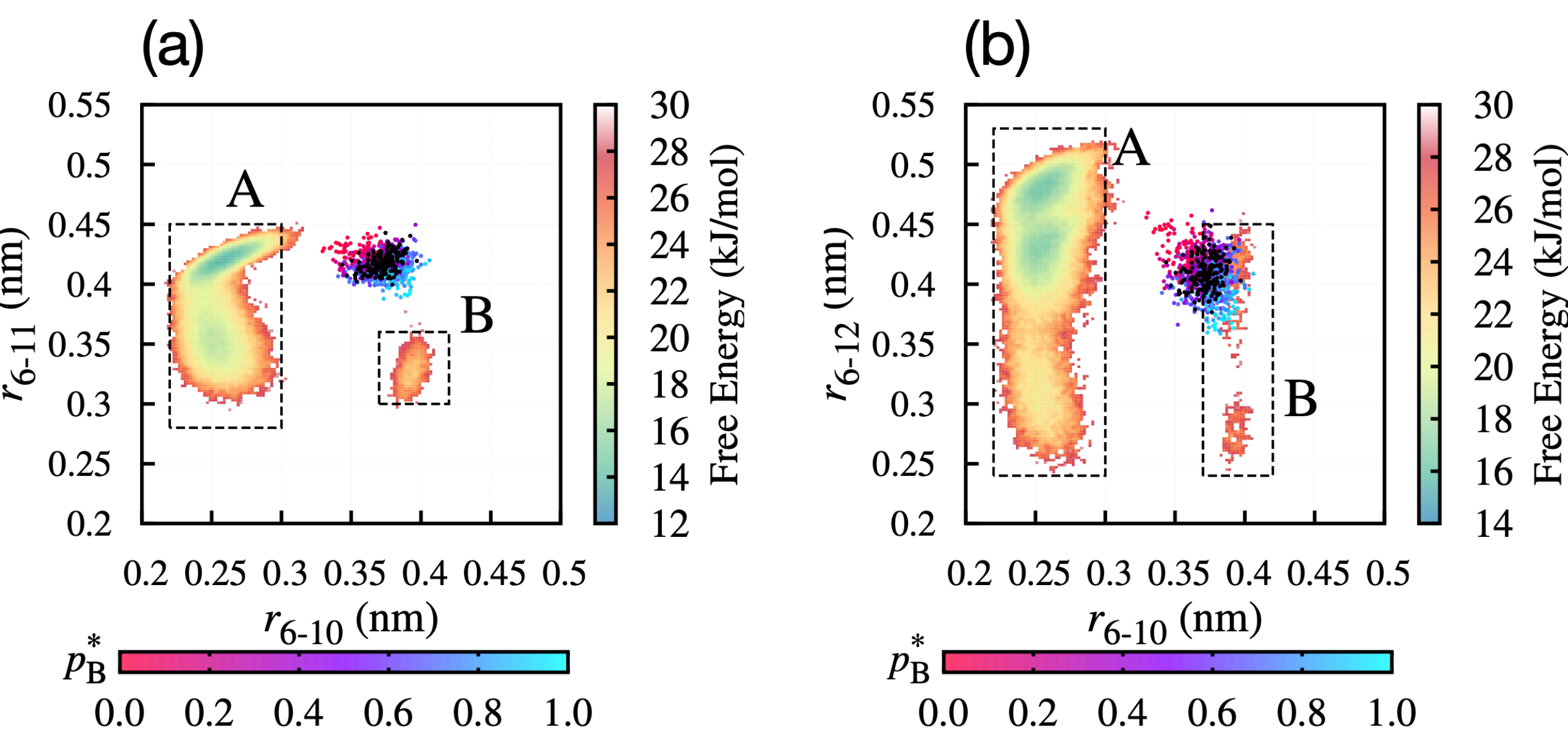}
\caption{
Two-dimensional FELs using ($r_{6-10}$, $r_{6-11}$) (a) and ($r_{6-10}$, $r_{6-12}$) (b).
The points represent sampled 1,000 shooting points (training 
 data set), which are colored by
 $p_\mathrm{B}^*$ values given in the bottom color bar.
The black boxes with dashed lines practically correspond to states A
 and B, characterized as 
[(0.22 nm, 0.28 nm) $\le$ ($r_{6-10}$, $r_{6-11}$) $\le$ (0.30 nm, 0.45
 nm)] in (a) and [(0.22 nm, 0.24 nm) $\le$ ($r_{6-10}$, $r_{6-12}$)
 $\le$ (0.30 nm, 0.53 nm)] in (b), and 
[(0.37 nm, 0.30 nm) $\le$ ($r_{6-10}$ 
 $r_{6-11}$) $\le$ (0.42nm, 0.36 nm)] in (a) and [(0.37 nm, 0.27 nm) $\le$ ($r_{6-10}$,
 $r_{6-12}$) $\le$ (0.42nm, 0.45 nm)] in (b), respectively.
Note that these boundaries are different from those in
 Fig.~\ref{fig:ala2}(b), which are used in the aimless shooting.
In addition, the points with $p_\mathrm{B}^* \sim 0.5$ $(0.45 \le p_\mathrm{B}^* \le 0.55)$ are marked in black dots.
}
\label{fig:2DPMF}
\end{figure*}
%%%%%%%%%%%%%%%%%%%%%%%%%%%%%%%%%%%%%%%%%%%%%%%%%%%%%%%%%%%%%%%%%%%%%%%%%%%%%%%%%%%%%%%%%%

\subsection{CV contribution using LIME and SHAP}
\label{sec:LIME_SHAP}

As detailed in Sec.~\ref{sec:XAI}, the
influence of each input variable on the neural network prediction can
be assessed through XAI methods, 
specifically LIME and Kernel SHAP.
These methods elucidate the manner in which
each feature contributes to the prediction using the additive feature attribution method.

We evaluated the feature contribution from
all of the data (2000 points) using LIME and SHAP.
The index dependence of the feature contribution in absolute value are
illustrated in Fig.~S1-S5 of the supplementary material
for each input variable case.
The comparison reveals that, for input variables with significant
contributions, LIME and SHAP provide nearly consistent
results. 
In contrast, for variables with smaller contributions, LIME
tends to yield output values that are more suppressed compared to SHAP,
particularly noticeable for indices 1-89 of 
interatomic distances.
Interestingly, LIME assesses that interatomic distancces involving atoms
1-5 are irrelevant to the target conformation change between the two
stable states, A and B.
This difference between LIME and SHAP can be attributed to the fact that
LIME quantifies the
contribution to prediction through linear regression, where the
correlation among input features is not taken into account, as mentioned
in Sec.~\ref{sec:XAI}.

Table~\ref{table:LIME_SHAP} presents the 
top five dominant contributions along with their absolute values for
each input variable case.
Remarkably, in cases (i), (ii), (iii), and (v), 
interatomic distances between atom 6 and atoms (10, 11, 12) occupy the top
two positions in both LIME and SHAP analyses.
As depicted in Fig.~\ref{fig:ala2}(a), 
these distances are related to the dominant two angles $\varphi$ and $\theta$,
making substantial contributions to the RC, $q$.
From an intuitive standpoint, one might expect the interatomic distance between atoms 6 and 18,
$r_{6-18}$, to be significant, given the consideration of 
hydrogen-bond formation.
However, our observations indicate that the extent of its importance is limited.
In fact, the variations in $r_{6-18}$ are connected to changes in the dihedral
angle $\psi$, a factor found to have a minor role in the investigated 
isomerization process.~\cite{bolhuis2000Reaction}

In case (iv), using 90 dihedral angles and 36 bond angles, the dihedral
angles $\theta$ and $\varphi$
emerge as the most dominant features.
This observation is consistent 
with a prior study 
that employed 90 dihedral angles 
as input variables.~\cite{kikutsuji2022Explaining}
However, 
the current results reveal 
an increased importance of $\theta$ relative to $\varphi$
compared to the previous findings.
This reversal can be attributed to the slight degradation in 
the learning performances associated with the inclusion of bond
angles, 
as discussed in Sec.~\ref{sec:committor_prediction}.
This influence is presumed to impact the outcomes of
LIME and SHAP analyses.

\subsection{FELs using interatomic distances}

The LIME and SHAP results consistently highlight that the interatomic distances
$r_{6-10}$, $r_{6-11}$, and $r_{6-12}$ might be the CVs that significantly
contribute to the RC, $q$.
Figure~\ref{fig:PMF} illustrates the FEL and corresponding committor
$p_\mathrm{B}^*$ distribution as a function of each CV, namely, 
$r_{6-10}$ (a), $r_{6-11}$ (b), and $r_{6-12}$ (c).
As evident in Fig.~\ref{fig:PMF}(a), the FEL using $r_{6-10}$ reveals
that 
the two stable states are 
distinguished by the free energy barrier. 
Furthermore, the position of the barrier, $r^*_{6-10}=0.36$ nm, approximately corresponds to $p_\mathrm{B}^*=0.5$,
representing the TS.
States A and B are practically delineated by specific ranges of the
interatomic distance $r_{6-10}$, namely 0.22 nm $\le$ $r_{6-10}$ $\le$
0.30 nm for state A and 0.37 nm $\le$ $r_{6-10}$ $\le$ 0.42 nm for state
B.
Since states A and B are originally defined in the FEL using $\varphi$
and $\psi$ as depicted in Fig.~\ref{fig:ala2}(b), 
the above thresholds expressed with $r_{6-10}$
may not 
exactly correspond to those defined by $\varphi$ and $\psi$.
In contrast, as depicted in Fig.~\ref{fig:PMF}(b) and (c), other
CVs, $r_{6-11}$ and $r_{6-12}$, exhibit less consistency in describing the FEL with the committor $p_\mathrm{B}^*$ distribution.
In particular, 
the free energy barriers are not distinctly observed, suggesting that these
variables alone are insufficient for describing the FEL.

Finally, it is of interest to explore
the two-dimensional FEL using possible combinations of the identified dominant 
CVs, namely ($r_{6-10}$, $r_{6-11}$) and ($r_{6-10}$, $r_{6-12}$), as illustrated in Fig.~\ref{fig:2DPMF}.
The combination of ($r_{6-11}$, $r_{6-12}$) was excluded
because the free energy barrier is insufficiently described using
$r_{6-11}$ and $r_{6-12}$, as observed in Fig.~\ref{fig:PMF}(b) and (c).
Figure~\ref{fig:2DPMF} illustrates the discrimination between
the two states, A and B, manifested by the separatrix
line composed of the data points with $p_\mathrm{B}^* \sim 0.5$,
indicative of the TS.
Note that the regions corresponding to states A and
B, as described in Fig.~\ref{fig:2DPMF}, may not precisely match those in
Fig.~\ref{fig:ala2}(b).
Mapping the rectangular region depicted in Fig.~\ref{fig:ala2}(b) using
($r_{6-10}$, $r_{6-11}$) and ($r_{6-10}$, $r_{6-12}$) would lead to a more
complex boundary than a simple rectangle.
As illustrated in Fig.~\ref{fig:2DPMF}(a), 
the two-dimensional FEL using $r_{6-10}$ and $r_{6-11}$ 
provides an alternative representation of FEL, complementing the FEL using
$\varphi$ and $\theta$ (see Fig.~\ref{fig:ala2}(c)).
Notably, the separatrix line indicates the
negative correlation between $r_{6-10}$ and $r_{6-11}$, suggesting 
that the barrier crossing from
state A to B requires 
an increase in $r_{6-10}$ and a decrease in $r_{6-11}$.
This behavior near the TS is consistent with the $r_{6-10}$ and $r_{6-11}$ dependence
of the committor distribution, as 
observed in Fig.~\ref{fig:PMF}(a) and (b).
A similar behavior near the TS is observed in the two-dimensional FEL using
$r_{6-10}$ and $r_{6-12}$ (see Fig.~\ref{fig:2DPMF}(b)), although the
separatrix line is not as clearly seen compared to Fig.~\ref{fig:2DPMF}(a).
Furthermore, the region corresponding to state B appears divided into 
two, with some overlap with the TS region.
These results suggest that $r_{6-11}$ is a more suitable CV than $r_{6-12}$
for describing the FEL alongside $r_{6-10}$.

\section{Conclusions}

Conventionally, 
the FEL concerning the conformational change in
alanine dipeptide has been effectively described using specific
main chain dihedral angles. 
However, our investigation recognized 
the need for 
a more detailed
characterization, through
the incorporation of more fundamental CVs such as interatomic distances
and bond angles, within the machine learning approach for the identifying RC.
By integrating XAI methods, specifically LIME and SHAP, into the machine learning predictions,
we assessed the importance of each input variable, and successfully
elucidated the specific interatomic distance, dominantly $r_\mathrm{6-10}$, in accord with
the committor $p_\mathrm{B}^*$ distribution.
The results manifest that constructing the FEL
using the identified interatomic distance differentiates
between the two stable states and clarifies the process of barrier
crossing through the TS.

Furthermore, our investigation into the two-dimensional FEL using the 
two dominant interatomic distances, $r_\mathrm{6-10}$ and
$r_\mathrm{6-11}$, provided insights into the conformational changes between states A and B. 
Specifically, the separatrix line, described by data points with
$p_\mathrm{B}^* \sim 0.5$, was clearly observed as the indicator of the TS.
This separatrix line aligns with the observations in the two-dimensional
FEL using two major dihedral angles, $\varphi$ and $\theta$, emphasizing 
the importance of considering specific interatomic distances in
capturing the target TS.
Nonetheless, utilizing interatomic distance as input variables implies a
reliance on pre-documented physical information. 
A potential future
perspective involves utilizing graph neural networks to autonomously
generate features based on the graph structure, thereby obtaining
RC without relying on the physics-informed variables.

Deciding between LIME and SHAP, or consistently employing
both, is an aspect worth considering.
The SHAP, developed as an additive feature
attribution method based on LIME, has the advantage of having a
theoretical basis in game theory. 
Since the prediction results are
distributed fairly among the features, it is expected that the
contribution of each feature can be quantified more accurately than LIME.
However, the computational cost of SHAP increases
exponentially with the number of features because it is necessary to
calculate the Shapley value for the number of combinations to be
considered. 
Given that the machine learning model used in this study is a
relatively simple regression on the sigmoidal function, which is a
monotonically increasing function, the variables with large
contributions identified by LIME and Kernel SHAP are considered to be consistent.

The optimization of hyperparameters for neural networks is another crucial
aspect to consider.
The selection of the numbers of hidden layers and nodes significantly
impacts 
the generalization performance of the neural network, thereby
influencing the
results obtained through XAI analyses. 
Our ongoing efforts are dedicated to exploring and implementing
optimizations in this direction.

\section*{Supplementary material}

The supplementary material include
definitions of dihedral angles, interatomic distances, and bond angles
(Tables~S1-S3), and 
feature contribution of each input variable using LIME and SHAP for each
input variable case
 (Figs.~S1-S5).

\begin{acknowledgments}
This work was supported by 
JSPS KAKENHI Grant-in-Aid 
Grant Nos.~\mbox{JP22H02595}, \mbox{JP22H02035}, \mbox{JP22H04542},
 \mbox{JP22K03550}, \mbox{JP23H02622}, \mbox{JP23K23303},
 \mbox{JP23K23858}, \mbox{JP23K27313}, and \mbox{JP24H01719}.
We acknowledge supports from
the Fugaku Supercomputing Project (Nos.~JPMXP1020230325 and JPMXP1020230327) and 
the Data-Driven Material Research Project (No.~\mbox{JPMXP1122714694})
from the
Ministry of Education, Culture, Sports, Science, and Technology.
The numerical calculations were performed at Research Center for 
Computational Science, Okazaki Research Facilities, National Institutes
of Natural Sciences (Projects: 24-IMS-C051 and 24-IMS-C198) and at the Cybermedia Center, Osaka University.
\end{acknowledgments}

%aipnum4-2.bst 2019-01-14 (MD) hand-edited version of apsrev4-1.bst

\section*{AUTHOR DECLARATIONS}

\section*{Conflict of Interest}
The authors have no conflicts to disclose.

\section*{Data availability statement}

The data that support the findings of this study are available from the
corresponding author upon reasonable request.

\clearpage
\widetext

\setcounter{equation}{0}
\setcounter{figure}{0}
\setcounter{table}{0}
\setcounter{page}{1}

\renewcommand{\theequation}{S.\arabic{equation}}
\renewcommand{\thefigure}{S\arabic{figure}}
\renewcommand{\thetable}{S\arabic{table}}
\renewcommand{\bibnumfmt}[1]{[S#1]}
\renewcommand{\citenumfont}[1]{S#1}

\noindent{\bf\Large Supplementary Material}
\vspace{5mm}
\begin{center}
\textbf{\large  
Unveiling interatomic distances influencing the reaction
 coordinates in alanine dipeptide isomerization:
 An explainable deep learning approach}
\\

\vspace{5mm}
 
{Kazushi Okada,$^1$ Takuma Kikutsuji,$^1$ Kei-ichi Okazaki,$^{2, 3}$
 Toshifumi Mori,$^{4, 5}$ Kang Kim,$^1$
and Nobuyuki Matubayasi$^1$}
\\

\vspace{5mm}

\noindent
\textit{
$^{1)}$Division of Chemical Engineering, Graduate School of Engineering Science, Osaka University, Osaka 560-8531, Japan}\\
\textit{
$^{2)}$Research Center for Computational Science, Institute for Molecular Science, Okazaki, Aichi 444-8585, Japan}\\
\textit{
$^{3)}$The Graduate University for Advanced Studies, Okazaki, Aichi 444-8585, Japan}\\
\textit{
$^{4)}$Institute for Materials Chemistry and Engineering, Kyushu University, Kasuga, Fukuoka 816-8580, Japan}\\
\textit{
$^{5)}$Interdisciplinary Graduate School of Engineering Sciences, Kyushu University, Kasuga, Fukuoka 816-8580, Japan}

\end{center}

%%%%%%%%%%%%%%%%%%%%%%%%%%%%%%%%%%%%%%%%%%%%%%%%%%%%%%%%%%%%%%%%%%%%%%%%%%%%%%%%%%%%%%%%%%
%%%%%%%%% Table S1 %%%%%%%%%%%%%%%%%%%%%%%%%%%%%%%%%%%%%%%%%%%%%%%%%%%%%%%%%%%%%%%%%%%%%%%%%
\begin{table}[h]
\caption{Definition of the dihedral angle index.
The atom index is represented in Fig.~1(a). 
Note that dihedral angles are used in cosine (index: 1-45)
 and sine (index: 46-90) forms.
}
\centering
      \begin{tabular}{l c c c}

\toprule
\midrule
      index & \multicolumn{3}{c}{index of atoms for dihedral angles} \\ \hline
      1-3 &  2 -  1 -  5 -  6 &  2 -  1 -  5 -  7  &  3 -  1 -  5 -  6 \\
      4-6 &  3 -  1 -  5 -  7 &  4 -  1 -  5 -  6  &  4 -  1 -  5 -  7 \\
      7-9 &  1 -  5 -  7 -  8 &  1 -  5 -  7 -  9  &  6 -  5 -  7 -  8 \\
      10-12 &   6 -  5 -  7 -  9 &  5 -  7 -  9 - 10  &  5 -  7 -  9 - 11 \\
      13-15 &   5 -  7 -  9 - 15 &  8 -  7 -  9 - 10  &  8 -  7 -  9 - 11 \\
      16-18 &   8 -  7 -  9 - 15 &  7 -  9 - 11 - 12  &  7 -  9 - 11 - 13 \\
      19-21 &   7 -  9 - 11 - 14 & 10 -  9 - 11 - 12  & 10 -  9 - 11 - 13 \\
      22-24 &  10 -  9 - 11 - 14 & 15 -  9 - 11 - 12  & 15 -  9 - 11 - 13 \\
      25-27 &  15 -  9 - 11 - 14 &  7 -  9 - 15 - 16  &  7 -  9 - 15 - 17 \\
      28-30 &  10 -  9 - 15 - 16 & 10 -  9 - 15 - 17  & 11 -  9 - 15 - 16 \\
      31-33 &  11 -  9 - 15 - 17 &  9 - 15 - 17 - 18  &  9 - 15 - 17 - 19 \\
      34-36 &  16 - 15 - 17 - 18 & 16 - 15 - 17 - 19  & 15 - 17 - 19 - 20 \\
      37-39 &  15 - 17 - 19 - 21 & 15 - 17 - 19 - 22  & 18 - 17 - 19 - 20 \\
      40-42 &  18 - 17 - 19 - 21 & 18 - 17 - 19 - 22  &  1 -  7 -  5 -  6 \\
      43-45 &   5 -  9 -  7 -  8 &  9 - 17 - 15 - 16  & 15 - 19 - 17 - 18 \\

\midrule
\bottomrule
      \end{tabular}
      \label{tab:dihedral} 
\end{table}
%%%%%%%%%%%%%%%%%%%%%%%%%%%%%%%%%%%%%%%%%%%%%%%%%%%%%%%%%%%%%%%%%%%%%%%%%%%%%%%%%%%%%%%%%%

\newpage
%%%%%%%%%%%%%%%%%%%%%%%%%%%%%%%%%%%%%%%%%%%%%%%%%%%%%%%%%%%%%%%%%%%%%%%%%%%%%%%%%%%%%%%%%%
%%%%%%%%% Table S2 %%%%%%%%%%%%%%%%%%%%%%%%%%%%%%%%%%%%%%%%%%%%%%%%%%%%%%%%%%%%%%%%%%%%%%%%%
\begin{table}[H]
\caption{Definition of the interatomic distance index.
The atom index is represented in Fig.~1(a). }
\centering
      \begin{tabular}{l c c c c c c c}

\toprule
\midrule
      index & \multicolumn{7}{c}{index of atoms for interatmoc distances} \\ \hline
      1-7     & 1-6   & 1-7   & 1-8   & 1-9   & 1-10  & 1-11  & 1-12 \\
      8-14    & 1-13  & 1-14  & 1-15  & 1-16  & 1-17  & 1-18  & 1-19 \\
      15-21   & 1-20  & 1-21  & 1-22  & 2-3   & 2-4   & 2-5   & 2-6  \\
      22-28   & 2-7   & 2-8   & 2-9   & 2-10  & 2-11  & 2-12  & 2-13 \\
      29-35   & 2-14  & 2-15  & 2-16  & 2-17  & 2-18  & 2-19  & 2-20 \\
      36-42   & 2-21  & 2-22  & 3-4   & 3-5   & 3-6   & 3-7   & 3-8  \\
      43-49   & 3-9   & 3-10  & 3-11  & 3-12  & 3-13  & 3-14  & 3-15 \\
      50-56   & 3-16  & 3-17  & 3-18  & 3-19  & 3-20  & 3-21  & 3-22 \\
      57-63   & 4-5   & 4-6   & 4-7   & 4-8   & 4-9   & 4-10  & 4-11 \\
      64-70   & 4-12  & 4-13  & 4-14  & 4-15  & 4-16  & 4-17  & 4-18 \\
      71-77   & 4-19  & 4-20  & 4-21  & 4-22  & 5-8   & 5-9   & 5-10 \\
      78-84   & 5-11  & 5-12  & 5-13  & 5-14  & 5-15  & 5-16  & 5-17 \\
      85-91   & 5-18  & 5-19  & 5-20  & 5-21  & 5-22  & 6-7   & 6-8  \\
      92-98   & 6-9   & 6-10  & 6-11  & 6-12  & 6-13  & 6-14  & 6-15 \\
      99-105  & 6-16  & 6-17  & 6-18  & 6-19  & 6-20  & 6-21  & 6-22 \\
      106-112 & 7-10  & 7-11  & 7-12  & 7-13  & 7-14  & 7-15  & 7-16 \\
      113-119 & 7-17  & 7-18  & 7-19  & 7-20  & 7-21  & 7-22  & 8-9  \\
      120-126 & 8-10  & 8-11  & 8-12  & 8-13  & 8-14  & 8-15  & 8-16 \\
      127-133 & 8-17  & 8-18  & 8-19  & 8-20  & 8-21  & 8-22  & 9-12 \\
      134-140 & 9-13  & 9-14  & 9-16  & 9-17  & 9-18  & 9-19  & 9-20 \\
      141-147 & 9-21  & 9-22  & 10-11 & 10-12 & 10-13 & 10-14 & 10-15\\
      148-154 & 10-16 & 10-17 & 10-18 & 10-19 & 10-20 & 10-21 & 10-22\\
      155-161 & 11-15 & 11-16 & 11-17 & 11-18 & 11-19 & 11-20 & 11-21\\
      162-168 & 11-22 & 12-13 & 12-14 & 12-15 & 12-16 & 12-17 & 12-18\\
      169-175 & 12-19 & 12-20 & 12-21 & 12-22 & 13-14 & 13-15 & 13-16\\
      176-182 & 13-17 & 13-18 & 13-19 & 13-20 & 13-21 & 13-22 & 14-15\\
      183-189 & 14-16 & 14-17 & 14-18 & 14-19 & 14-20 & 14-21 & 14-22\\
      190-196 & 15-18 & 15-19 & 15-20 & 15-21 & 15-22 & 16-17 & 16-18\\
      197-203 & 16-19 & 16-20 & 16-21 & 16-22 & 17-20 & 17-21 & 17-22\\
      204-210 & 18-19 & 18-20 & 18-21 & 18-22 & 20-21 & 20-22 & 21-22\\

\midrule
\bottomrule
      \end{tabular}
      \label{tab:distance} 
\end{table}
%%%%%%%%%%%%%%%%%%%%%%%%%%%%%%%%%%%%%%%%%%%%%%%%%%%%%%%%%%%%%%%%%%%%%%%%%%%%%%%%%%%%%%%%%%

%%%%%%%%%%%%%%%%%%%%%%%%%%%%%%%%%%%%%%%%%%%%%%%%%%%%%%%%%%%%%%%%%%%%%%%%%%%%%%%%%%%%%%%%%%
%%%%%%%%% Table S3 %%%%%%%%%%%%%%%%%%%%%%%%%%%%%%%%%%%%%%%%%%%%%%%%%%%%%%%%%%%%%%%%%%%%%%%%%
\begin{table}[H]
\caption{Definition of the bond angle index.
The atom index is represented in Fig.~1(a). 
Note that the bond angles are used in cosine forms.
}
\centering
      \begin{tabular}{l c c c c}

\toprule
\midrule
      index & \multicolumn{4}{c}{index of atoms for bond angles} \\ \hline
      1-4     & 2-1-3     & 2-1-4     & 2-1-5     & 3-1-4   \\
      5-8     & 3-1-5     & 4-1-5     & 1-5-6     & 1-5-7   \\
      9-12    & 6-5-7     & 5-7-8     & 5-7-9     & 8-7-9   \\
      13-16   & 7-9-10    & 7-9-11    & 7-9-15    & 10-9-11 \\
      17-20   & 10-9-15   & 11-9-15   & 9-11-12   & 9-11-13 \\
      21-24   & 9-11-14   & 12-11-13  & 12-11-14  & 13-11-14\\
      25-28   & 9-15-16   & 9-15-17   & 16-15-17  & 15-17-18\\
      29-32   & 15-17-19  & 18-17-19  & 17-19-20  & 17-19-21\\
      33-36   & 17-19-22  & 20-19-21  & 20-19-22  & 21-19-22\\

\midrule
\bottomrule
      \end{tabular}
      \label{tab:bond} 
\end{table}
%%%%%%%%%%%%%%%%%%%%%%%%%%%%%%%%%%%%%%%%%%%%%%%%%%%%%%%%%%%%%%%%%%%%%%%%%%%%%%%%%%%%%%%%%%

\newpage
%%%%%%%%%%%%%%%%%%%%%%%%%%%%%%%%%%%%%%%%%%%%%%%%%%%%%%%%%%%%%%%%%%%%%%%%%%%%%%%%%%%%%%%%%%
%%%%%%%%% Fig. S1 %%%%%%%%%%%%%%%%%%%%%%%%%%%%%%%%%%%%%%%%%%%%%%%%%%%%%%%%%%%%%%%%%%%%%%%%%
\begin{figure}[H]
\centering
\includegraphics[scale=0.5]{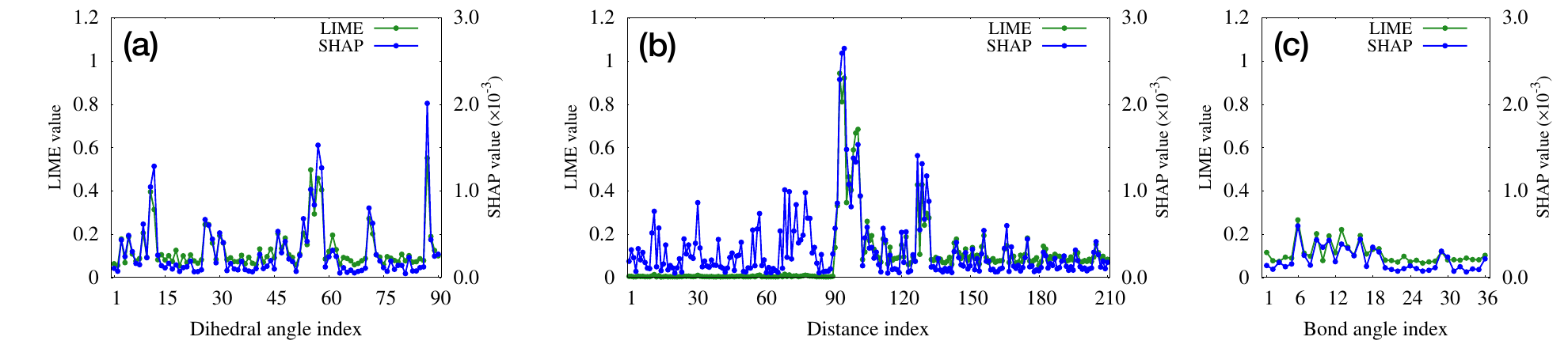}
\caption{
Dependence of feature contribution on CV index, represented in absolute
 values, 
 obtained through LIME (green) and SHAP (blue) for the neural network
 predictions using (i) dihedral angles + interatomic distances +
 bond angles.
Panels (a), (b), and (c) display the index dependence of 
90 dihedral angles, 210 diatomic distances, and 36 bond angles, respectively.
}
\end{figure}
%%%%%%%%%%%%%%%%%%%%%%%%%%%%%%%%%%%%%%%%%%%%%%%%%%%%%%%%%%%%%%%%%%%%%%%%%%%%%%%%%%%%%%%%%%

%%%%%%%%%%%%%%%%%%%%%%%%%%%%%%%%%%%%%%%%%%%%%%%%%%%%%%%%%%%%%%%%%%%%%%%%%%%%%%%%%%%%%%%%%%
%%%%%%%%% Fig. S2 %%%%%%%%%%%%%%%%%%%%%%%%%%%%%%%%%%%%%%%%%%%%%%%%%%%%%%%%%%%%%%%%%%%%%%%%%
\begin{figure}[H]
\centering
\includegraphics[scale=0.5]{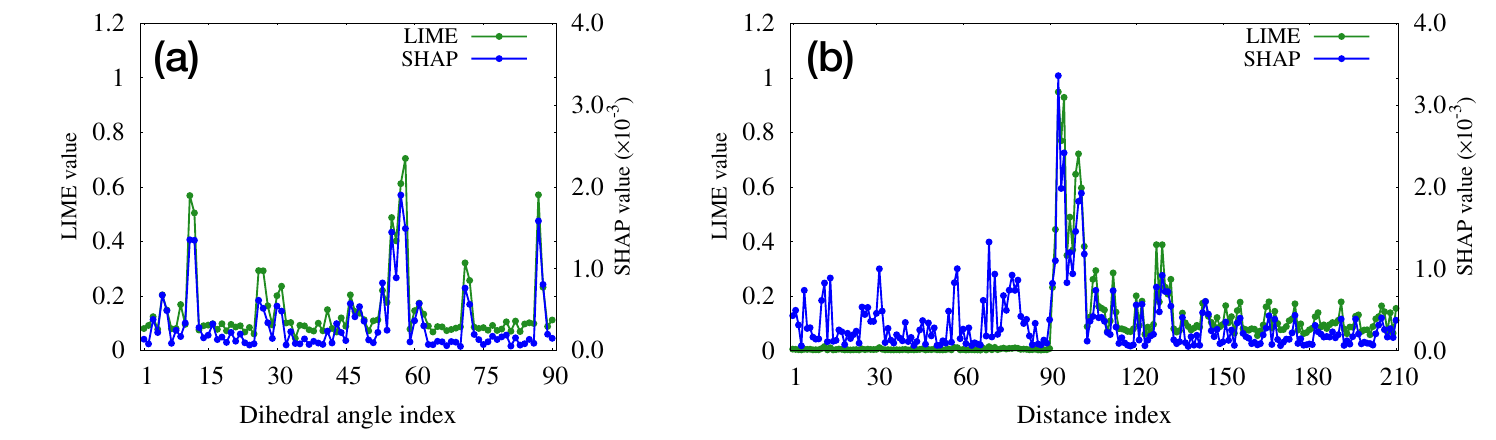}
\caption{
Dependence of feature contribution on CV index, represented in absolute
 values, 
 obtained through LIME (green) and SHAP (blue) for the neural network
 predictions using (ii) 
dihedral angles + interatomic distances.
Panels (a) and (b) display the index dependence of 
90 dihedral angles and 210 diatomic distances, respectively.
}
\end{figure}
%%%%%%%%%%%%%%%%%%%%%%%%%%%%%%%%%%%%%%%%%%%%%%%%%%%%%%%%%%%%%%%%%%%%%%%%%%%%%%%%%%%%%%%%%%

%%%%%%%%%%%%%%%%%%%%%%%%%%%%%%%%%%%%%%%%%%%%%%%%%%%%%%%%%%%%%%%%%%%%%%%%%%%%%%%%%%%%%%%%%%
%%%%%%%%% Fig. S3 %%%%%%%%%%%%%%%%%%%%%%%%%%%%%%%%%%%%%%%%%%%%%%%%%%%%%%%%%%%%%%%%%%%%%%%%%
\begin{figure}[H]
\centering
\includegraphics[scale=0.5]{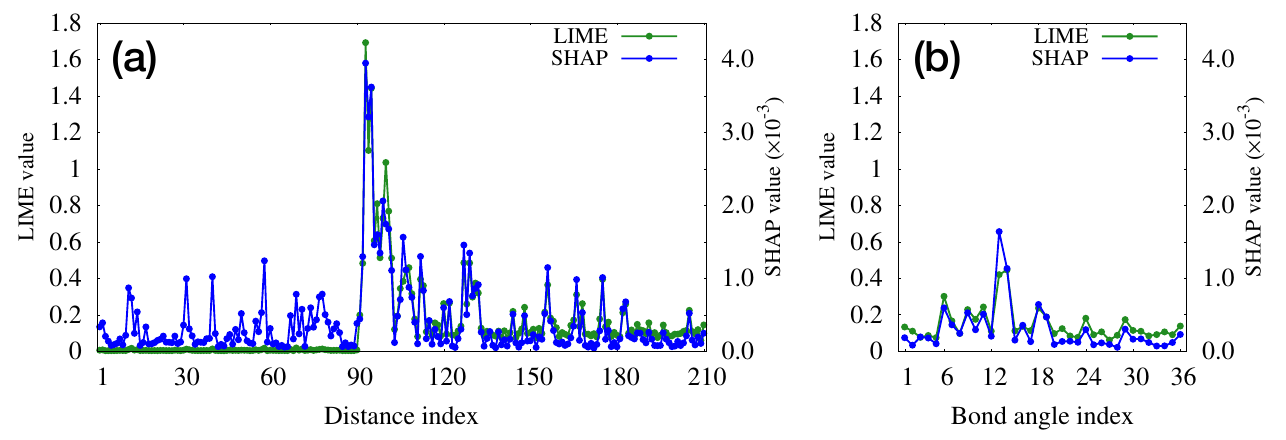}
\caption{
Dependence of feature contribution on CV index, represented in absolute
 values, 
 obtained through LIME (green) and SHAP (blue) for the neural network
 predictions using (iii) 
interatomic distances + bond angles.
Panels (a) and (b) display the index dependence of 
210 diatomic distances and 36 bond angles, respectively.
}
\end{figure}
%%%%%%%%%%%%%%%%%%%%%%%%%%%%%%%%%%%%%%%%%%%%%%%%%%%%%%%%%%%%%%%%%%%%%%%%%%%%%%%%%%%%%%%%%%

%%%%%%%%%%%%%%%%%%%%%%%%%%%%%%%%%%%%%%%%%%%%%%%%%%%%%%%%%%%%%%%%%%%%%%%%%%%%%%%%%%%%%%%%%%
%%%%%%%%% Fig. S4 %%%%%%%%%%%%%%%%%%%%%%%%%%%%%%%%%%%%%%%%%%%%%%%%%%%%%%%%%%%%%%%%%%%%%%%%%
\begin{figure}[H]
\centering
\includegraphics[scale=0.5]{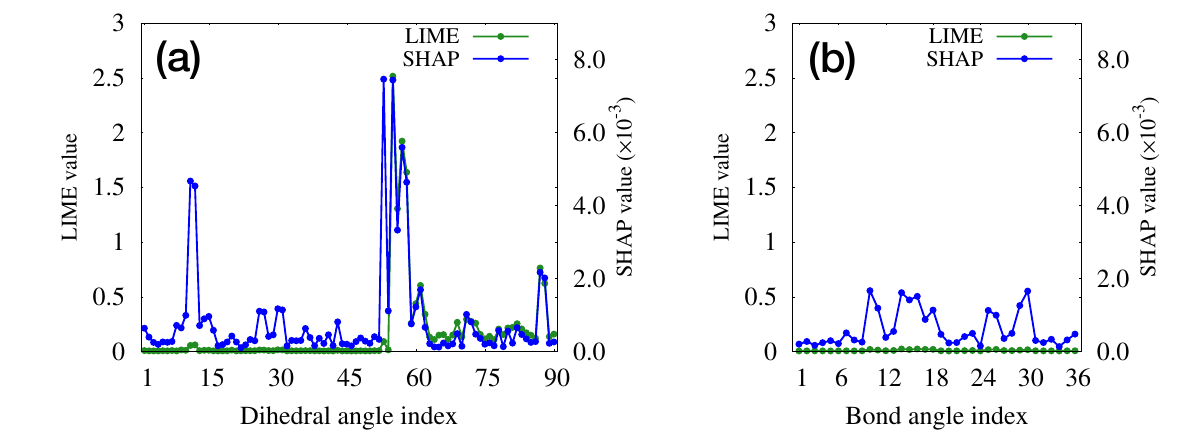}
\caption{
Dependence of feature contribution on CV index, represented in absolute
 values, 
 obtained through LIME (green) and SHAP (blue) for the neural network
 predictions using (iv) 
dihedral angles + bond angles.
Panels (a) and (b) display the index dependence of 
90 dihedral angles and 36 bond angles, respectively.
}
\end{figure}
%%%%%%%%%%%%%%%%%%%%%%%%%%%%%%%%%%%%%%%%%%%%%%%%%%%%%%%%%%%%%%%%%%%%%%%%%%%%%%%%%%%%%%%%%%

%%%%%%%%%%%%%%%%%%%%%%%%%%%%%%%%%%%%%%%%%%%%%%%%%%%%%%%%%%%%%%%%%%%%%%%%%%%%%%%%%%%%%%%%%%
%%%%%%%%% Fig. S5 %%%%%%%%%%%%%%%%%%%%%%%%%%%%%%%%%%%%%%%%%%%%%%%%%%%%%%%%%%%%%%%%%%%%%%%%%
\begin{figure}[H]
\centering
\includegraphics[width=0.5\textwidth]{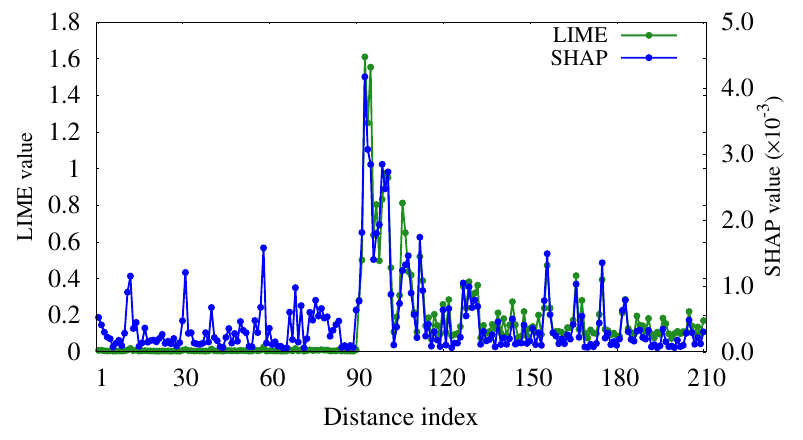}
\caption{
Dependence of feature contribution on CV index, represented in absolute
o values, 
 obtained through LIME (green) and SHAP (blue) for the neural network
 predictions using 
210 interatomic distances.
}
\end{figure}
%%%%%%%%%%%%%%%%%%%%%%%%%%%%%%%%%%%%%%%%%%%%%%%%%%%%%%%%%%%%%%%%%%%%%%%%%%%%%%%%%%%%%%%%%%

\end{document}